\documentstyle[prb,aps,preprint,epsf]{revtex}
\tighten
\begin{document}
\title{Re-entrant spin susceptibility of a superconducting grain}
\author{
        A. Di Lorenzo$^{(1)}$, Rosario Fazio$^{(1,2)}$, F.W.J.
        Hekking$^{(3,4)}$, G. Falci$^{(1,2)}$, A. Mastellone$^{(1,2)}$,
        and G. Giaquinta$^{(1,2)}$}
\address{
$^{(1)}$Dip. di Metodologie Fisiche e Chimiche (DMFCI),
        Universit\`a di Catania,\\
        viale A. Doria 6, 95125 Catania, Italy\\
$^{(2)}$Istituto Nazionale per la Fisica della Materia (INFM),
        Unit\'a di Catania, Italy\\
$^{(3)}$Theoretische Physik III, Ruhr-Universit\"at Bochum, 44780 Bochum,
        Germany\\
$^{(4)}$CNRS-CRTBT \& Universit\'e Joseph Fourier, 38042 Grenoble Cedex 9, France
        }
\date{\today}
\maketitle

\begin{abstract}
We study the spin susceptibility $\chi$ of a small, isolated
superconducting grain.  Due to the interplay between parity effects
and pairing correlations, the dependence of $\chi$ on temperature $T$
is qualitatively different from the standard BCS result valid in the
bulk limit.  If the number of electrons on the grain is odd, $\chi$
shows a {\em re-entrant} behavior as a function of temperature.  This
behavior persists even in the case of ultrasmall grains where the
mean level spacing is much larger than the BCS gap.  If the number of
electrons is even, $\chi (T)$ is exponentially small at low
temperatures.
\end{abstract}

\pacs{PACS numbers: 74.20.Fg, 7323.Hk, 74.80.Bj}

\narrowtext
By now it is well-known that the properties of an isolated, mesoscopic
superconducting grain are quite different from those of a bulk
sample~\cite{curacao}.  First of all, since such a grain carries a
fixed number, $N$, of electrons, its behavior depends strongly on whether
$N$ is even or odd.  Second, fluctuation effects become
important as the size of the grain decreases.  The interplay
between parity and fluctuation effects crucially depends on the ratio
$\delta / \Delta _{0}$ of two characteristic energies: the mean level
spacing $\delta$ and the bulk superconducting gap $\Delta _{0}$.  As
long as the grain is not too small, $\delta \ll \Delta _{0}$, the
fluctuation region $\Delta T$ around the critical temperature $T_c$ is
narrow, $\Delta T/T_{c} \sim \sqrt{\delta/\Delta _{0}} \ll 1$, and the
mean field description of superconductivity is appropriate.  Parity
effects~\cite{Averin92,Tuominen92} appear at temperatures lower than a
crossover temperature $T_{\rm eff} \simeq \Delta_{0}/ \ln \sqrt{8\pi
\Delta_{0}^{2}/ \delta ^{2}}$ which, in the
experiments~\cite{Tuominen92}, is typically of the order of $10-30 \%$
of $T_{c}$.  The dependence of $T_{\rm eff}$ on $\Delta _{0}$ signals
that the even-odd asymmetry is a collective effect due to pairing
correlations.  As the size of the grain is decreased, fluctuations
start to smear the superconducting transition~\cite{Muehlschlegel72}.
The finite level spacing suppresses the BCS gap in a parity-dependent
way~\cite{vonDelft96,Smith96}.  When $\delta$ becomes of the order of
$\Delta_{0}$, $\Delta T \sim T_{c}$ and the BCS description of
superconductivity breaks down even at zero
temperature~\cite{Anderson59}.  The regime $\delta \gtrsim \Delta_{0}$
is dominated by strong pairing 
fluctuations~\cite{Richardson64,Matveev97,Mastellone98,Berger98,Braun98,Dukelsky99}.

The present-day interest in ultrasmall superconducting grains was
triggered by the experiments of Ralph, Black and
Tinkham~\cite{Ralph_Black95&96}, who were able to contact a single,
nanometer-sized Al grain with current and voltage probes. They
obtained tunneling spectra that revealed the presence of a
parity-dependent spectroscopic gap, larger than the average level spacing, which
could be driven to zero by an applied magnetic field.  
In this Letter we propose to measure the temperature dependence of a
thermodynamic quantity -- the spin susceptibility $\chi$ -- as a means
to detect {\em both} parity effects {\em and} pairing correlations.
As we will show below, pairing correlations
give rise to a specific temperature dependence of thermodynamic
quantities. This enables a more quantitative investigation of
fluctuation effects~\cite{footnote2}.

\begin{figure}
\vspace{2cm}
\centerline{{\epsfxsize=9cm\epsfysize=6.3cm\epsfbox{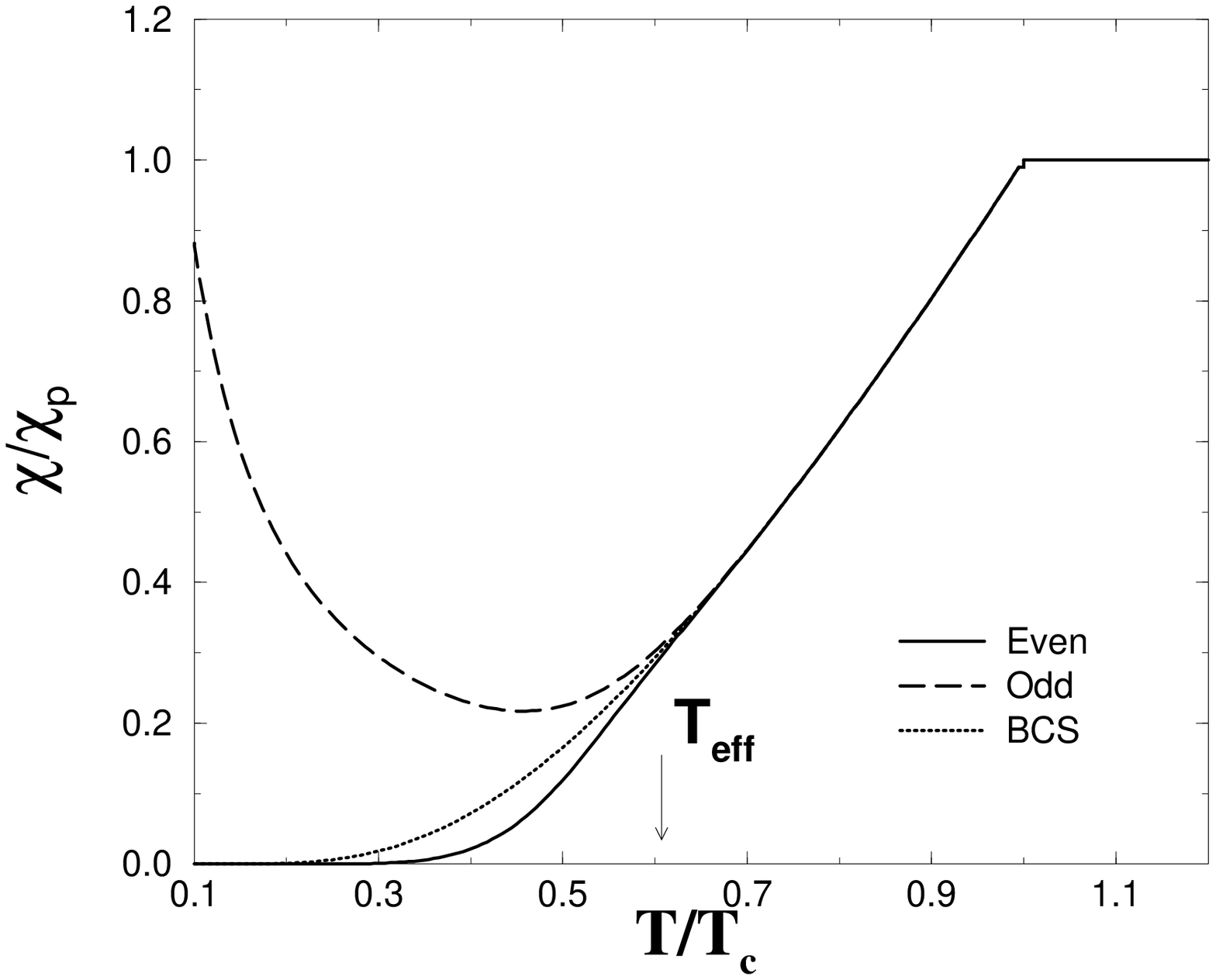}}}
\vspace{0.2cm}
\caption{$\Delta _{0} \gg \delta $ -- Spin susceptibility as a function
of temperature $T/T_{c}$ for an odd (upper curve) and even (lower
curve) grain, respectively.  The BCS result (middle curve) is reported
for comparison.  The susceptibility is normalized to its bulk high
temperature value $\chi_{P} = 2 \mu^2_B /\delta $.  We used
$\delta/\Delta _{0} =0.1$.}

\label{fig1}
\end{figure}

\begin{figure}
\centerline{{\epsfxsize=9cm\epsfysize=7.2cm\epsfbox{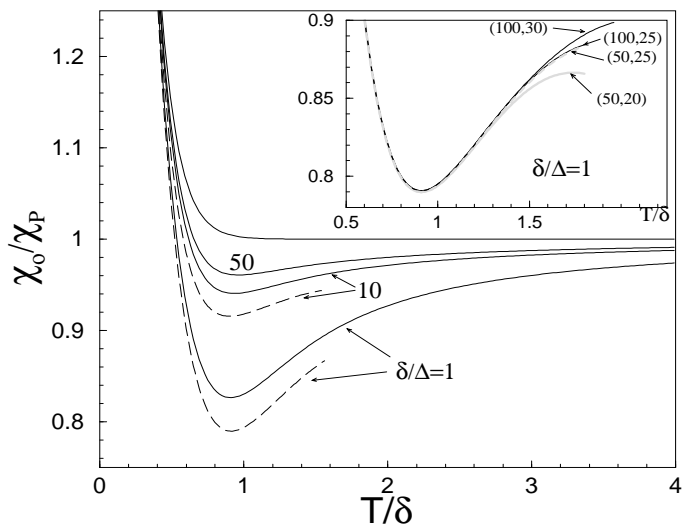}}}
\vspace{0.2cm}
\caption{$\Delta _{0} \le \delta $ -- Reentrant spin susceptibility as
a function of temperature for an odd grain in the static path
approximation (solid lines); the top-most curve without re-entrance
is the non-interacting limit. The dashed lines are obtained by an
exact canonical calculation (see text), using $N=100$ electrons and
a large enough maximum excitation energy ($\Lambda \sim 40 \delta$).
In the inset the dependence of the susceptibility as a function of $N$
and $\Lambda$. Curves are labelled with $(N,\Lambda)$.
}
\label{fig2}
\end{figure}

\begin{figure}
\centerline{{\epsfxsize=9cm\epsfysize=7.2cm\epsfbox{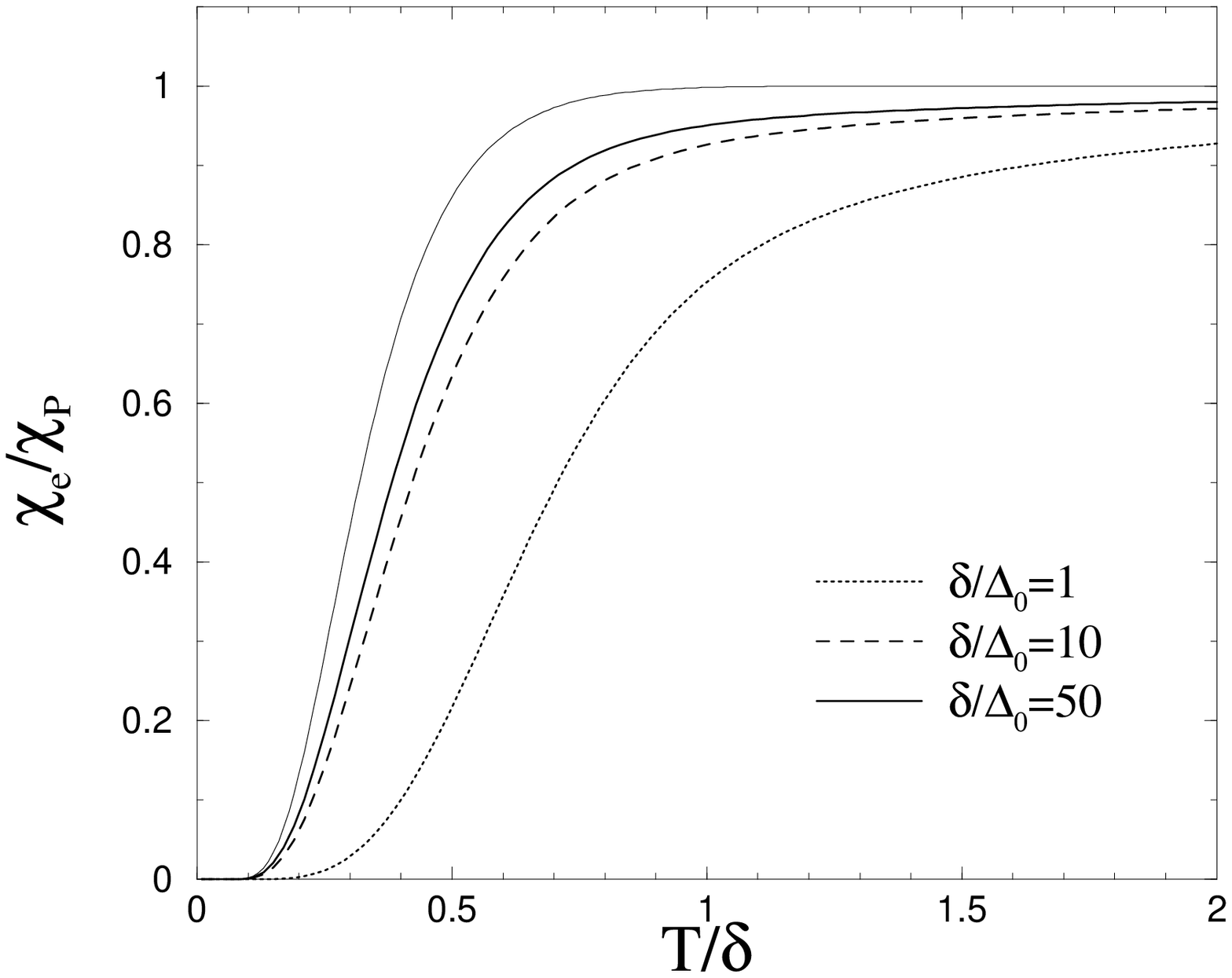}}}
\vspace{0.2cm}
\caption{$\Delta _{0} \le \delta $ -- Spin susceptibility as a function
of temperature for an even grain. As in the previous figure, the thin solid line
is the non-interacting limit.
}
\label{fig3}
\end{figure}

Spin paramagnetism of small particles has been considered in the
past~\cite{Perenboom81} and very recently parity effects in the
susceptibility were measured for an ensemble of small, normal metallic
grains~\cite{Volotikin96}.  Spin susceptibility is very sensitive to
BCS pairing as well.  Yosida~\cite{Yosida58} showed that, due to the
opening of the superconducting gap, $\chi $ vanishes at zero
temperature~\cite{footnote1}.  We will show that the combined effect
of parity and pairing introduces {\em qualitatively} new features in
the temperature dependence of $\chi$.  Most interestingly, these effects
might be observed even in the regime $\delta \gtrsim \Delta_{0}$.  The
results of our work are summarized in Figs.~\ref{fig1} -- \ref{fig3},
where we plot $\chi$ as a function of temperature $T$ for
odd and even parity.  In particular we want to emphasize that the odd
susceptibility shows a {\em re-entrant} behavior as a function of $T$
{\em for any value of the ratio} $\delta / \Delta _{0}$.  This
re-entrance is absent in normal metallic grains; it is a genuine feature
of the interplay between pairing correlations and parity effects.

The BCS pairing Hamiltonian for a small grain can be written
as~\cite{vonDelft96,Richardson64,Matveev97}
\begin{equation}
        {\cal H} = \sum_{{n, \sigma=\pm}} (\epsilon_n -\sigma \mu _B H)\,
        c_{n,\sigma}^{\dagger}  c_{n,\sigma} - \lambda \delta
        \sum_{m,n}B_{m}^{\dagger} B_{n} ,
\label{hamiltonian}
\end{equation}
where $B_{m}^{\dagger} =c_{m,+}^{\dagger}c_{m,-}^{\dagger} $.  The
indices $m$, $n$ label the single particle energy levels with
energy $\epsilon_m$ and annihilation operator $c_{m,\sigma}$.  The
quantum number $\sigma =\pm$ labels time-reversed equally spaced states
(with an average spacing $\sim \delta \equiv 1/\nu _{0}$, where $\nu _{0}$ 
is the density of states at the Fermi energy).  The external magnetic 
field $H$ couples to the electrons via the Zeeman term, $\mu_B$ is the Bohr 
magneton. We put the $g$-factor equal to two, ignoring any spin-orbit effects
(see Ref.~\onlinecite{Braun97}).  At the low magnetic fields of
interest here, we can neglect the orbital contribution to magnetic
energy, as it is smaller than the Zeeman energy by a factor $\sim
(k_{F}r) (Hr^2/\Phi _{0})$ ($r$ is the size of the grain and
$\Phi_{0}$ the flux quantum).  Finally $\lambda$ is the
dimensionless BCS coupling constant.  Since the Hamiltonian contains only pairing 
terms, an
electron in a singly occupied level cannot interact with the other
electrons.

The spin susceptibility of a grain with an even (e) or an odd (o)
number $N$ of electrons is defined as
\begin{equation}
        \chi_{e/o} (T)= - \left. \frac{\partial ^{2}
        {\cal F}_{e/o}(T,H)}{\partial H^{2}}
        \right|_{H=0} ,
\label{chi}
\end{equation}
where ${\cal F}_{e/o} = - T \ln Z_{e/o}$ is the free energy of
the grain and the partition function $Z(T,N)$ should be evaluated in
the canonical ensemble. We will perform the calculation
with the help of a parity projection technique~\cite{Janko94,Golubev94} and by
means of exact canonical methods based on Richardson's solution~\cite{Richardson64}.
The grand partition function reads
\begin{equation}
        Z_{e/o}(T,\mu) = (1/2)\sum_{N=0}^{\infty} e^{ \mu N/T}
        [1 \pm e^{i \pi N}] Z(T,N)
        \equiv  (1/2)\left(Z_+ \pm Z_-\right) .
\label{partition}
\end{equation}
The partition function $Z_{+}$ is the usual grand partition function
at temperature $T$ and chemical potential $\mu_{+}=\mu$.  The grand partition
function $Z_{-}$ describes an auxiliary ensemble at temperature $T$
and chemical potential $\mu_{-}=\mu + i\pi T$; it is a formal tool, necessary
to include parity effects.  The chemical potential $\mu$ will be
placed between the topmost occupied level and the lowest unoccupied
level in the even case, while it will be at the singly occupied level
in the odd case.  Since we are interested in the evaluation
of fluctuations effects, it is convenient to express the grand
partition functions $Z_{\pm}$ using the path integral formulation of
superconductivity~\cite{Muehlschlegel72,reviews,Golubev94},
\begin{equation}
        Z_{\pm}= Z_\pm ^0
\frac{
        \int
        {{\cal D}^2 \Delta
        \exp\left\{\int \limits_{0}^{\beta}
        {d\tau \left[
        {\rm Tr} \ln (1 - \hat{G}_{\pm}^0\hat{\Delta})-
        \frac{\mid \Delta \mid^2}{\lambda \delta}
        \right]}\right\}}
}
{
        \int
        {\cal D}^2\Delta
        \exp\left\{- \int \limits_{0}^{\beta}
        {d\tau \frac{\mid \Delta \mid^2}{\lambda \delta}}\right\}
} .
\label{hubz}
\end{equation}
Here, $\beta =1/T$ and $Z_\pm ^0$ is the partition function for
non-interacting electrons.  The matrix Green function 
$\hat{G}_{\pm}^{0}$ is given by
$        
	\hat{G}_\pm ^{(0)} (\epsilon_{\nu}) =
        \left[ (i\omega_{\nu} + \mu_B H) \sigma ^{(0)} -
        (\epsilon_{n} - \mu_{\pm} ) \sigma ^{(z)}
        \right]^{-1} ,
$
where $\omega_{\nu}$ is a fermionic Matsubara frequency,
$\sigma ^{(i)}$ ($i=x,y,z$) are the Pauli
matrices, and $\sigma^{(0)}$ is the identity.  Finally, the matrix
$\hat{\Delta}$ is given by
$
        \hat{\Delta} =
        (\Delta/2)(\sigma ^x +i \sigma ^y) + \rm{h.c.}
$
A direct calculation of the partition function~(\ref{hubz}) is
impossible in general.  Below, we first discuss two limiting cases
which are tractable analytically: $\delta /\Delta _{0} \ll 1$ and
$\delta /\Delta _{0} \gg 1$.  Then we present the complete temperature
dependence of the spin susceptibility evaluating Eq.~(\ref{hubz})
numerically for arbitrary values of $\delta /\Delta _{0}$,
with the help of the static path approximation~\cite{Muehlschlegel72}.

\underline{Large grains ($ \; \Delta _{0} \gg \delta$) }.  In this
limit it is sufficient to evaluate the partition function in a saddle
point approximation, since fluctuations will not contribute
significantly~\cite{Averin94}.  As a result we find
\begin{equation}
        \chi _{e/o}
        =
        \frac{\mu _B^2}{2T}\sum _n
        \frac{Z_+ \cosh ^{-2} \frac{E_{+,n}}{2T}
        \mp Z_-  \sinh ^{-2} \frac{E_{-,n}}{2T} }{Z_+\pm Z_-} ,
\label{susc}
\end{equation}
where $E_{\pm,n} = \sqrt{\epsilon_n^2 + \Delta_{\pm}}$. The saddle
point values of $\Delta_{\pm}$ are found from the equations
\begin{equation}
\frac{1}{\lambda}
        =
        \sum \limits _{n,\sigma} \frac{\delta}{4E_{\pm ,n}}
        \mbox{th}^{\pm 1}\left(\frac{E_{\pm,n} -
        \sigma \mu_BH }{2T} \right) .
\label{sad}
\end{equation}
The partition functions for the two ensembles are
\begin{eqnarray}
        Z_{\pm}
        =
        \exp
        \left\{\sum \limits _{n,\sigma} \left[\ln 2
        \left\{ \begin{array}{c} \mbox{ch}      \\ \mbox{sh}
        \end{array}     \right\}
        {\frac{E^{\sigma}_{\pm,n}}{2T}}
        -  \frac{\xi _n }{2T}\right] -
        \frac{\Delta _{\pm}^2}{\lambda \delta T} \right\},
\end{eqnarray}
where $E^{\sigma}_{\pm,n} = E_{\pm,n} -\sigma \mu_B H$ and $\xi_n
=\epsilon _n -\mu$.

At low temperatures $T \ll \Delta_{0}$, the ratio $Z_{-}/Z_{+}$ can be
calculated easily; one finds $Z_{-}/Z_{+} \simeq 1 - \sqrt{8\pi
T\Delta_{0}/\delta ^{2}} \exp{(-\beta \Delta _{0})}$.  Parity effects are
important if this ratio is $\sim 1$, {\em i.e.} at temperatures
$T < T_{\rm eff}$.  At temperatures $T_{\rm eff} \ll T \ll
\Delta _{0}$, parity effects can be ignored and the spin susceptibility is
found to decrease exponentially, as in the BCS case,
\begin{equation}
        \chi _{e/o} \sim \frac{2 \mu _B^2}{\delta} 
		\sqrt{\frac{2\pi \Delta _{0}}{T}}
        e^{-\beta \Delta _{0}}.
                \label{chiBCS}
\end{equation}
For $T \ll T_{\rm eff}$, Eq.~(\ref{susc}) can be approximated as
$$
        \chi_e   \, \simeq \, \frac {8\pi\mu _{B}^{2}\Delta _{0}}{\delta^2}
        e^{-2 \beta \Delta _{0}} \; ; \quad
                \chi_o   \,  \simeq \, \frac {\mu _{B}^{2}}{T}
        .
$$
We see that $\chi _{e}$ remains exponentially small, like in the BCS
case (\ref{chiBCS}), but with an exponent
$-2 \beta \Delta _{0}$
rather than $-\beta \Delta _{0}$.  This reflects the fact that
excitations are actually created in pairs.  In odd grains, the
unpaired spin gives rise to an extra paramagnetic (Curie-like)
contribution to the spin susceptibility.  As a result $\chi_o$ will
show a {\em re-entrant effect} at low temperatures (see Fig.~\ref{fig1}).
Although the re-entrant behavior is essentially a single electron effect,
we stress that it can be detected experimentally using granular systems with
many well-separated grains (to avoid collective effects due to
tunneling).  Such systems contain even grains as well, but their
susceptibility is exponentially small at the temperatures of interest;
thus their contribution to the response of the system will be
negligible.

\underline{Ultrasmall grains ($ \; \Delta _{0}\ll \delta$) }.
A reduction of the grain size leads to a suppression of the gap
$\Delta $.  For ultrasmall grains with $\Delta _0 \ll \delta$,
the mean field approximation gives $\Delta = 0$:
the grain behaves as a normal metal.
The non-interacting, parity-dependent spin
susceptibility can be found from Eq.~(\ref{susc}), see the topmost curves
in Figs.~\ref{fig2} and \ref{fig3}.
Note in particular the monotonous dependence
of $\chi _{o}$ on $T$.
The temperature scale at which parity effects
appear is set by the average level spacing.
If $T \gg \delta$, parity effects are
exponentially small and $\chi _{e/o} (T) \simeq
\chi_{P} [1 \mp (2T/\delta) \exp (-\pi ^2T/\delta)]$.
In the opposite limit
$T \ll \delta$, $\chi _{e}$ is exponentially
small, $\chi _{e}(T) \simeq (8 \mu _{B}^{2}/T)e^{-\beta \delta}$, as we need to
excite an
electron out of the topmost, doubly occupied single particle level to
magnetize the grain. For an odd grain, $\chi _{o}(T) \simeq \mu _{B}^{2}/T$ at 
$T \ll \delta$:
the topmost level is occupied by a single electron that gives a
Curie-like contribution.

The saddle point approach entirely ignores the fact that
the fluctuation region $\Delta T$ around
$T_c$ grows as the size of the grains is reduced.
Due to the presence of fluctuations, the behavior of small grains will be
different in a distinct way from normal metallic grains.
In the limit $T \gg \delta$, both fluctuation and parity effects are small;
it therefore suffices to consider the fluctuation correction
$\delta \chi _{\rm fluc}$ to $\chi_P$, evaluating $Z_{+}$,
Eq.~(\ref{hubz}), in Gaussian approximation.  As a
result, we find $\delta \chi _{\rm fluc}/\chi_P \simeq -
\delta/2T\ln(T/\Delta_0)$; hence
$\chi _{e/o} (T) \simeq
\chi_{P} [1 \mp (2T/\delta) \exp (-\pi ^2T/\delta) - \delta/2T\ln(T/\Delta_0)]$.
Superconducting correlations {\em suppress} the susceptibility;
due to its algebraic dependence on $T$ this suppression is stronger than
the parity correction at temperatures
$T \gtrsim \delta \ln \ln \delta/\Delta_{0}$.
In the opposite limit,
$T \ll \delta$, fluctuations are strong and the Gaussian approximation
fails.  However, the susceptibility can still be obtained analytically
by considering a few levels close to the Fermi energy with a
renormalized pairing interaction $\tilde{\lambda} = 1 /\ln
(\delta/\Delta_0)$~\cite{Matveev97,Berger98}.  Consider first a grain
with an even number of electrons. It costs an energy
$\sim \delta +\delta /\ln (\delta/\Delta_0)$ to excite an electron from the
topmost, doubly occupied level to the lowest unoccupied level. Correspondingly,
the leading temperature dependence of
the spin susceptibility is
\begin{equation}
        \chi_e (T)
        \simeq
        8\frac{\mu _B^2}{T} e^{-\beta \delta(1 + \ln^{-1} (\delta/\Delta_0)) }
        + {\cal O}(e^{-2\beta \delta}).
\end{equation}
The even susceptibility is exponentially small, like in the case of a normal 
metallic grain, but
with an exponent
$-\beta \delta (1 + \tilde{\lambda})$, rather than $-\beta \delta$.
Similarly, we find the spin susceptibility for a grain with an odd
number of electrons
\begin{equation}
        \chi_o (T)
        \simeq
        \frac{\mu _B^2}{T} [1+8e^{-\beta \delta(2 + \ln^{-1}
        (\delta/\Delta_0)) }].
\end{equation}
The paramagnetic contribution from the single spin dominates at all
temperatures below $\delta$. Compared to the case of a normal metallic grain,
the odd susceptibility is
non-monotonous: upon lowering temperature, $\chi _{o}$ first
 decreases due to superconducting fluctuations; at temperatures
 $T \sim \delta$ a re-entrant
behavior sets in which persists down to the lowest temperatures.

\underline{Reentrant susceptibility.}
The various limiting cases discussed so far provide
evidence for the appearance of an anomaly in the spin susceptibility
$\chi _{o}$. For large grains ($\Delta \gg \delta$) the mean field
approximation, Eq.(\ref{susc}), leads to the re-entrant behaviour
shown in Fig.~\ref{fig1}. We will show that this is a unique signature of
pairing correlations which is present even in ultrasmall grains.
To this end we study the complete temperature dependence of $\chi _{e/o}$
for arbitrary values of the ratio $\delta/\Delta _{0}$.

The physics of re-entrant susceptibility
can be grasped by evaluating Eq.~(\ref{hubz}), in  the static path
approximation~\cite{Muehlschlegel72}.
This amounts in retaining only the static fluctuations (beyond the
Gaussian approximation) in the path integral.
In Figs.~\ref{fig2} and \ref{fig3} we show the results of this
calculation for the odd and the even cases respectively.
The re-entrance in the odd case is visible even in systems with
a ratio $\delta/\Delta_{0} \sim 50$ (!) and provides the signature
of the existance of pairing correlations in the ultrasmall regime.
Frequency dependent fluctuations will further enhance the size of the
reentrant effect. The results are plotted for a system of $N=200$ electrons
at half filling and the BCS coupling is chosen to fix the ratio $\delta/\Delta$
($\lambda \sim 0.1-0.2$). The merit of the static path approximation combined 
with the analytic
analysis in the limiting cases is that it allows to
obtain a coherent quantitative physical picture in the whole temperature
range.

As a final check of our results we computed $\chi_o$ using the
exact solution of Ref.~\cite{Richardson64}. The result is presented in
Fig.\ref{fig2} (dashed lines).
As expected, the re-entrant effect is slightly larger ($\sim 15$\%).
In order to obtain this result we considered all the different
states with excitation energy up to a cutoff $\Lambda \sim 40 \delta$ 
for a system with
$N \le 100$ electrons. In the inset we show the scaling analysis for
different $N$ and different energy cutoffs. This analysis
becomes more and more difficult upon increasing temperature because of
the exponential increase of the number of excited states needed.

In this Letter we proposed the study of  the spin susceptibility of a
metallic grain as a very sensitive probe to detect superconducting
correlations. For grains in the nanometer size regime the odd spin
susceptibility is a unique signature of pairing. 
In grains of dimensions of the order of few nanometers as those studied
in Refs.~\onlinecite{Ralph_Black95&96} the reentrance should be of the
order of $10-20$\% of the Pauli value and it could be measured using the
techinque used in Ref.~\onlinecite{Volotikin96}.

{\bf Acknowledgments} We thank I. Aleiner, I. Beloborodov,
A. Larkin, and B. M\"uhlschlegel for useful discussions.
We acknowledge the financial support of European Community (Contract
FMRX-CT-97-0143), SFB237 of the Deutsche Forschungsgemeinschaft and
INFM-PRA-QTMD.

\begin{references}
\bibitem{footnote2} It is quite difficult to obtain information
        concerning fluctuation effects from tunneling spectra in the limit
        $\delta \gg \Delta _{0}$, as it is necessary to look for small
        quantitative changes compared to the spectrum of a normal metallic 
	grain.
\end{references}
\end{document}